\documentstyle[prl,aps,preprint,tighten,graphics,psfrag,floats]{revtex}
\begin{document}                % INITIALIZE - DONT CHANGE

\title{A Model of Convergent Extension in Animal Morphogenesis}
\author{Mark Zajac\thanks{mzajac@krypton.helios.nd.edu}, Gerald
L. Jones and James A. Glazier }
\address{ University of Notre Dame,
Department of Physics Notre Dame, IN 46556 }
\date{\today}
\maketitle

\begin{abstract}

In this paper we argue that the pattern of cell movements in the
morphogenetic process known as convergent extension can be understood
as an energy minimization process, provided the cell-cell adhesive
energy has a certain type of anisotropy. This single simple property
suffices to cause the cell elongation, alignment, and intercalation of
a cellular array that are the characteristics of convergent
extension. We describe the type of anisotropy required. We show that
the final aspect ratio of the array of cells is independent of the
initial configuration and of the degree of cell elongation. We find
how it depends on the anisotropy.

\end{abstract}
 % 87.18.Hf 87.18.La 87.18.-h 87.17.Aa 87.18.Bb}

In the development of the animal embryo great changes of form
(morphogenesis) take place \cite{wolpert:development}, especially
during gastrulation when axial structures are formed by extensive cell
rearrangement.  During these rearrangements groups of cells move
coherently over distances very large compared to cell dimensions.
This process has been extensively investigated by experiments,
particularly on embryos of the frog, {\it Xenopus laevis}, and
particularly by R.E.  Keller and his collaborators (see
\cite{keller:les.houches} for a brief review and extensive
references).

\begin{figure}
\psfrag{a}{a}
\psfrag{b}{b}
\psfrag{c}{c}
\centering
\resizebox{12cm}{!}{%
  \includegraphics{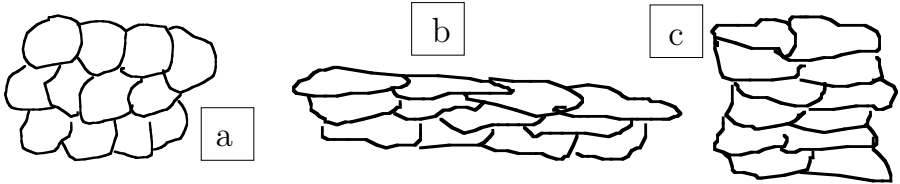}
}
\caption{Intercalation.  Isodiametric cells (a) become elongated and
aligned (b)while {\it simultaneously} intercalating (c) so that an
array of cells extends at right angles to the direction of cell
motion.}
\label{intercalation}
\end{figure}

One characteristic and widespread type of rearrangement, ``convergent
extension'', occurs, for example, in the development of axial
structure such as precursors to the vertebrate spinal column.  Here an
active group of cells undergoes a threefold process.  The individual
cells, originally roughly isodiametric (Fig.\ \ref{intercalation}a),
{\bf elongate} and their axes of elongation become {\bf aligned}.  If
these were the only motions the final configuration would be as in
Fig.\ \ref{intercalation}b.  But at the same time, though on a
somewhat slower time scale, the cells {\bf intercalate} between each
other.  The intercalation is in the direction of alignment so that the
number of cells in that direction decreases while the number of cells
in directions perpendicular to the alignment increases producing a
final configuration as in Fig.\ \ref{intercalation}c.  The elongation
increases the overall length of the group of cells in the direction of
alignment and decreases the length in orthogonal directions (since the
volume stays roughly constant).  Intercalation does the reverse but
dominates, so that the axis of net {\bf extension} of the group of
cells is at right angles to the axis of individual cell {\bf
elongation}.  Here we argue that the important aspects of convergent
extension will result from a tendency of the active cells to minimize
their total energy, provided that they interact with a non-uniform
surface (adhesive) energy satisfying certain conditions.  We also
develop a mean field theory of this process.

Minimum energy principles have been used to explain cell rearrangement
since Steinberg's \cite{steinberg:reconstruction} suggestion that
differential cell adhesion plus cell motility can account for cell
sorting patterns in mixtures of two or more cell types (see
\cite{graner:i} for a review and extensive references to the
literature).  Goel and Lieth \cite{goel:anisotropic} have considered
cell sorting for a simple geometrical model in the presence of
anisotropic surface adhesion between cells of fixed shape.  Cell
sorting, driven by energy minimization, has also been the subject of
many computer simulations \cite{graner:simulation,glazier:simulation}.
Drasdo, Kree, and McCaskill \cite{drasdo:monte.carlo} have simulated
cell sorting with anisotropic surface adhesion.  Anisotropic surface
adhesion has not, so far, been used to explain the convergent
extension of a homogeneous group of cells.  We do not model here the
dynamics of convergent extension.  We assume, as in
\cite{steinberg:reconstruction} and \cite{goel:anisotropic}, that cell
motility will allow the system to explore its possible configurations
and that, as a strongly dissipative system, it will evolve towards the
configuration of minimum energy.

In the embryo convergent extension usually takes place in an
asymmetric environment where the inactive cells bounding the active
region are not the same on all sides of that region.  In this case,
the extension, and its orientation, may be determined by the
interactions at the boundaries which ``channel'' the active cells,
rather than being an intrinsic collective property of the group of
active cells.  Under these experimental circumstances the boundaries
strongly influence active cell movements.  Indeed, in the physical
model of Weliky {\it et.al.}  \cite{weliky:notochord}, convergent
extension results only if active cells at the boundaries parallel to
the elongation, behave differently from those at the boundaries
perpendicular to the elongation.

A subsequent and elegant experiment by Shih and Keller
\cite{shih:organizer}, however, strongly suggests that, in addition,
the active cells have a strong intrinsic collective mechanism driving
their convergent extension.  In these experiments a layer (essentially
a monolayer) of active cells was excised from a frog embryo, at a
stage before convergent extension had begun, and cultured on a uniform
surface in a medium which allowed their normal development.
Subsequently the layer showed strong convergent extension in the plane
of the substrate - and this in the absence any plausible lateral
anisotropy either in the substrate or in the culture medium.  This
behavior thus appears to be an example of ``broken symmetry'' so well
known in condensed matter physics, and asks for an explanation based
on collective behavior induced by cell-cell interactions.

To explain this behavior by energy minimization we assume
that cell-cell interactions take place through surface adhesion, which
can be characterized by an energy per unit contact area.  We assume
that the cell rearrangements take place with negligible cell division
and little change in cell volume, as is observed in the later stages
of the above experiment.  There is no clear understanding in
the literature of the trigger for the cell elongation which initiates
the convergent extension, and our model does not provide this.  Our
main assumption is that the adhesive energy of the contact surface
between two cells will depend on how that surface is oriented relative
to the axes of elongation of the two cells.  This would be so,
for example, were the surface density of adhesive binding sites to be
different on the long side of a cell (parallel to the axis of
elongation) from that on the short sides (perpendicular to the axis of
elongation).  We can find in the literature no compelling evidence
either for or against this assumption.  We argue here that a specific
form of this assumption is a sufficient cause of the elongation,
alignment, and intercalation resulting in convergent extension.

\begin{figure}
\psfrag{a}{a}
\psfrag{b}{b}
\centering
\resizebox{12cm}{!}{%
  \includegraphics{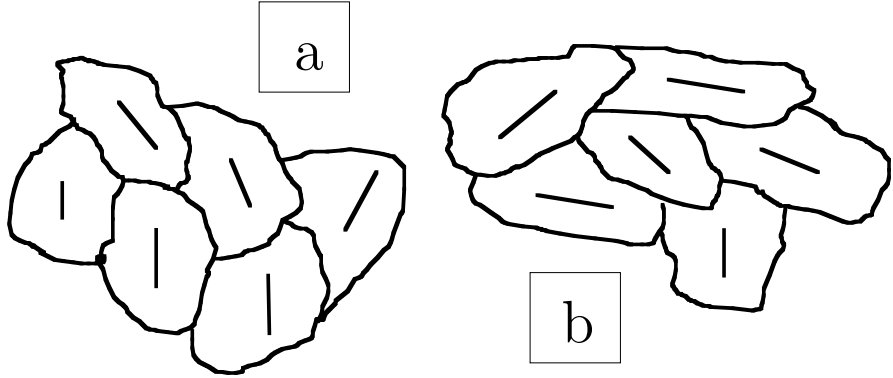}
}
\caption{Cell alignment.  For an ordered array (a) most cell
attachments are either end to end or side to side while a disordered
array (b) exhibits significant binding between poles and lateral
surfaces.}
\label{cartoon}
\end{figure}

We give here a two-dimensional version of our proposal since the
convergent extension takes place in the plane of the substrate and the
height of the cells does not seem to play an important role.  Hence we
consider a collection of two dimensional cells of (nearly) the same
fixed area.  We first want to find the conditions which favor
alignment.  We assume a compact array of cells large enough that array
surface effects are (for this argument) negligible.  Fig.\
\ref{cartoon}a is a cartoon of a few elongated cells in such a large
ordered array of cells and Fig.\ \ref{cartoon}b is for a disordered
array.  Suppose that we can roughly distinguish, for each cell, two
long sides (parallel to the axis of elongation) and two short sides
(perpendicular).  Figure\ \ref{cartoon}a shows that in the ordered
array the cell-cell contact surfaces are, for the most part, either
roughly parallel to the common axes of alignment or roughly
perpendicular to that axis.  We shall term these long-long ($ll$) or
short-short ($ss$) contacts since they occur, primarily, at contacts
between a pair of long sides or a pair of short sides.  The disordered
array of Fig.\ \ref{cartoon}b has many contact surfaces that make
intermediate angles with the now different axes of adjacent cells.  We
term these long-short ($ls$) contacts since they tend to occur whith
the contact surfaces between a long side of one cell and a short side
of a neighbor.  If the energy density (per unit length) of the $ls$
contacts is enough larger than those of $ll$ and $ss$ contacts then
the ordered array will have the lower energy per cell (we have assumed
that the array is large enough that we can neglect the effect of the
array boundaries on the bulk ordering).  More quantitatively, let $l$
and $s$ be the average long and short side lengths of each cell,
which, for the moment, we take as fixed.  Suppose that all cell-cell
contacts can be characterized as $ll$, $ss$, or $ls$ and that the
total lengths of each type in the array are $L_{ll}$, $L_{ss}$, and
$L_{ls}$.  In a large array of $N$ cells we have $2Nl=2L_{ll}+L_{ls}$
and $2Ns=2L_{ss}+L_{ls}$ (again neglecting array boundaries where
cells do not contact other cells).  Since $N$, $l$ and $s$ are fixed
these equations provide two constraints between the three contact
lengths.  We assume that three energy densities ($J_{ll}$, $J_{ss}$,
and $J_{ls}$) are adequate to characterize the interactions at the
various surfaces.  Then the bulk energy of an array due to cell-cell
interactions is
\begin{eqnarray}
E & = & L_{ll}J_{ll}+L_{ss}J_{ss}+L_{ls}J_{ls} \label{energy}    \\
  & = & (2Nl-L_{ls})J_{ll}/2
    + (2Ns-L_{ls})J_{ss}/2
    + L_{ls}J_{ls}                       \nonumber  \\
  & = & (J_{ls}-J_{ll}/2-J_{ss}/2)L_{ls}
    + N(lJ_{ll}+sJ_{ss}) \mbox{.}         \nonumber
\end{eqnarray}
This energy is an increasing function of $L_{ls}$ if the ordering
condition:
\begin{equation}
\gamma_{ls} = J_{ls} - (J_{ll} + J_{ss})/2 > 0\mbox{,}
\label{tension}
\end{equation}
is satisfied.  In this event ordered arrays ($L_{ls}=0$) will have
lower bulk energies than disordered ($L_{ls}>0$) arrays.  Note that
condition\ (\ref{tension}) is just that the $ls$ surface tension
$\gamma_{ls}$ be positive.

The above argument is exact if the cells are assumed (unrealistically)
to be identical rectangles arranged in arbitrary tesselations of the
plane and is similar to that used in \cite{goel:anisotropic} in the
cell sorting problem.  For realistic cells it is a crude but plausible
representation of the assumed anisotropy of the surface adhesion.

The case where the adhesive energy density of a two cell contact is
the product of a factor from each cell is interesting.  For example,
where a variation in the density of the binding sites on the cell
surface causes the variation in energy.  If we make the natural
assumption that the density of adhesive bonds is proportional to the
product of the density of binding sites on the cell surfaces in
contact, then we would have in the above model $J_{ll}=-j_lj_l$,
$J_{ss}=-j_sj_s$, and $J_{ls}=-j_lj_s$, where the sign is chosen make
all $J<0$ when all $j>0$.  This choice satisfies the ordering
condition Eq.\ (\ref{energy}) whenever $j_l$ and $j_s$ are positive
and are not equal.

In addition to Eq.\ (\ref{energy}) let us suppose that the $ll$ energy
density is lower than the $ss$ energy density.
\begin{equation}
J_{ll} < J_{ss} < 0\;(\mbox{or}\;j_l > j_s > 0)\mbox{.}
\label{order}
\end{equation}
Now the energy Eq.\ (\ref{energy}) of the array can be reduced by
increasing the cell long side length $l$ and decreasing the short
side length $s$ causing, or at least favoring, elongation of the
cells.  At equilibrium these surface effects will presumably be
balanced by internal cellular forces opposing further elongation.

We can also argue that Eq.\ (\ref{order}) will produce intercalation
in the direction of elongation.  Consider the effect of the boundary
on a finite array of N cells.  Suppose that boundary cells have no
adhesive energy with the culture medium.  Then the expression Eq.\
(\ref{energy}) underestimates the array energy because it assumes all
cell surfaces are in contact with other cell surfaces and so
overestimates the contact lengths $L_{ll}$ and $L_{ss}$.  From Eq.\
(\ref{energy}) we should subtract the (negative) adhesive energy that
is not present at the contacts between the boundary cells and the
surrounding medium.  Figure\ \ref{intercalation} shows arrays of twelve
elongated cells.  In Fig.\ \ref{intercalation}c the array extension is
at right angles to the cell elongation and in Fig.\
\ref{intercalation}b it is along the cell elongation.  Clearly in
Fig.\ \ref{intercalation}c the boundary contacts are primarily through
short cell sides whereas in Fig.\ \ref{intercalation}b they are
primarily though long cell sides.  Since the long sides have lower
(more negative) energy than the short, the energy (corrected for
boundaries) of the configuration shown in Fig.\ \ref{intercalation}b
is higher than that of Fig.\ \ref{intercalation}c.  Thus if we start
with any compact initial array of unelongated cells we expect cell
motility and energy minimization to produce configuration of type
Fig.\ \ref{intercalation}c by cell elongation, alignment, and intercalation
parallel to the alignment.  In order for these processes to produce
net extension in the direction perpendicular to alignment the effects
of intercalation must dominate those of elongation.  For a rectangular
array of a large number of rectangular cells a direct calculation
shows that this will be so, independent of the degree of elongation,
and that the ratio of the array dimensions in the directions
perpendicular and parallel to the elongation is $J_{ll}/J_{ss}$.  We
derive these results more generally below.

\begin{figure}
\psfrag{A}{a}
\psfrag{B}{b}
\psfrag{a}{$\hat{\bf a}$}
\psfrag{n}{$\hat{\bf n}$}
\centering
\resizebox{12cm}{!}{%
  \includegraphics{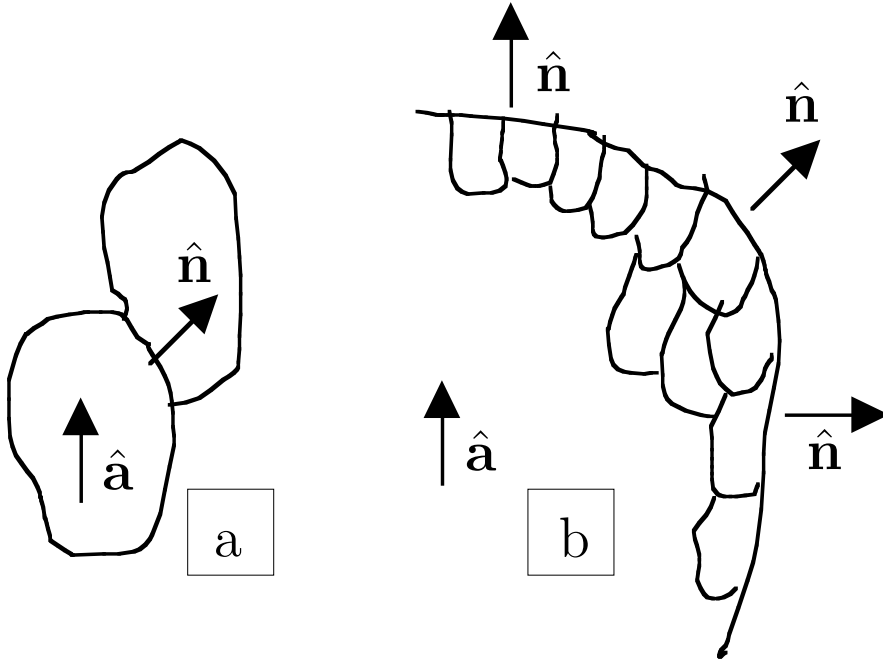}
}
\caption{Anisotropic binding.  Adhesive energy at the point of contact
between cells (a) is assumed to depend on $(\hat{\bf n} \cdot \hat{\bf
a})^2$ where $\hat{\bf n}$ is the local unit normal while $\hat{\bf
a}$ gives alignment, assumed common to all cells.  At an interface
with uniformly inert surroundings (b), missing adhesive energy will
vary with the orientation of the surface cells, relative to the
boundary.}
\label{anisotropy}
\end{figure}

We can make the above arguments concerning surface effects somewhat
more realistic and quantitative by the following mean field type of
modeling.  We assume that we have a large array of $N$ elongated and
aligned cells.  The total energy of the array is the bulk energy due
to cell-cell interactions plus the surface correction for the absence
of cells outside the boundary.  The bulk energy is proportional to
$N$, or equivalently, the array area $A$, so we write it as $\lambda
A$, where $\lambda$ is the (negative) bulk energy per unit area in the
aligned array.  To model the anisotropic cell-cell interaction we
assume that $J$ depends on the angle between the direction of
alignment, specified by the unit vector $\hat{\bf a}$, and the unit
vector $\hat{\bf n}$ normal to the contact segment between the cells
(see Fig.\ \ref{anisotropy}a).  More explicitly, we assume that $J(\hat{\bf
n} \cdot \hat{\bf a})$ is negative, an even function (since $\hat{\bf
a}$, $-\hat{\bf a}$ and $\hat{\bf n}$, $-\hat{\bf n}$ specify the same
physical situations), and is minimum at $\hat{\bf n} \cdot \hat{\bf
a}=0$ (so that $ll$ interactions have the lowest energy).  Figure\
\ref{anisotropy}b shows part of a finite array of vertically aligned
cells and their boundary with an external medium with which we assume
they have no adhesive energy.  To get the energy of the finite array
we must subtract from the bulk energy half the energy the boundary
cells would have had with cells external to the array had the boundary
been absent.  Half, since adhesive energy is shared between two cells.
So:
\begin{equation}
E=\lambda A
 - \frac{1}{2}
   \oint J(\hat{\bf n} \cdot \hat{\bf a})\,\mbox{d}l\mbox{,}
\label{functional}
\end{equation}
where the integral is taken around a closed boundary.  We want to
minimize this over all closed boundaries enclosing the same area $A$.
Alternatively we can interpret $\lambda$ as a Lagrange multiplier and
find the extrema of Eq.\ (\ref{functional}) over all closed curves at
fixed $\lambda$.  To do this parameterize the curves as ${\bf r}(u)$
with $0 \leq u \leq 1$, and ${\bf r}(0)={\bf r}(1)$.  Then, since
$\mbox{d}l = (\dot{x}^2 + \dot{y}^2)^{1/2}\,\mbox{d}u$ (where
$\dot{\bf r}=d{\bf r}/du$), while $(\hat{\bf n} \cdot \hat{\bf
a})=(a_y\dot{x} - a_x\dot{y})/(\dot{x}^2 + \dot{y}^2)^{1/2}$ and
$A=\int^{1}_{0} y\dot{x}\,\mbox{d}u$ we can write the energy as
$\int^{1}_{0} {\cal L}({\bf r},\dot{\bf r})\,\mbox{d}u$ with ${\cal
L}({\bf r},\dot{\bf r})=\lambda y \dot{x} - J(\hat{\bf n} \cdot
\hat{\bf a})(\dot{x}^2 + \dot{y}^2)^{1/2}/2$.  The extremal curves are
solutions of the usual Euler-Lagrange equations for ${\cal L}$ and are
degenerate with respect to translations in the $x$-$y$ plane.  This
degeneracy gives rise to two first integrals and two constants of
integration (which we choose to be zero), which fix the position of
the extremal curve.  The integrated equations have the form
\begin{equation}
    2 \lambda {\bf r}
  = \hat{\bf a} J^{\prime}(\hat{\bf n} \cdot \hat{\bf a})
  + \hat{\bf n}[J(\hat{\bf n} \cdot \hat{\bf a})
              - ( \hat{\bf n} \cdot \hat{\bf a} )
                J^{\prime}(\hat{\bf n} \cdot \hat{\bf a})]\mbox{,}
\label{trajectory}
\end{equation}
where $J^{\prime}$ is the derivative of $J$.  The dot product of Eq.\
\ref{trajectory} with $\hat{\bf r}$ gives the Wulff condition
\cite{wulff:construction}.  Equation\ (\ref{trajectory}) is two
coupled first order differential equations whose solutions depend on
the particular choice of the function $J$.  We have not been able to
find complete analytic solutions for any interesting choice of $J$ but
some properties of the solutions can be found.  First we note that for
$\hat{\bf a} = 0$, or equivalently $J = constant$, the solution
is a circle of radius $J/(2 \lambda)$.  Secondly, the turning points
of any solution curve are where $d({\bf r} \cdot {\bf r})/du = 0$.
Now:
\begin{equation}
\lambda \frac{d}{du}\!({\bf r} \cdot {\bf r})
  = 2 \lambda (\dot{\bf r} \cdot {\bf r})
  = (\dot{\bf r} \cdot \hat{\bf a})
    J^{\prime}( \hat{\bf n} \cdot \hat{\bf a} )\mbox{,}
\label{turning}
\end{equation}
where we have used Eq.\ (\ref{trajectory}) and that $\dot{\bf r} \cdot
\hat{\bf n}=0$ for any curve.  From Eq.\ (\ref{turning}) we find two
types of turning points.  1) At $\dot{\bf r} \cdot \hat{\bf a}=0$,
that is, where the boundary is perpendicular to the alignment so that
$\hat{\bf n}=\pm \hat{\bf a}$.  For any simple closed curve this
condition will be satisfied at two points on the curve.  Because $J$
is even and $J^{\prime}$ is odd we have from (\ref{trajectory}) that
these two points lie at $\pm J(1)/(2 \lambda)$ on the line through the
origin and parallel to $\hat{\bf a}$.  2) At $\hat{\bf n} \cdot
\hat{\bf a}=0$ where $J^{\prime}=0$ and the boundary is parallel to
the alignment.  For these (\ref{trajectory}) shows that $2\lambda {\bf
r}=\hat{\bf n} J(0)$, thus there are turning points at $\pm J(0)/(2
\lambda)$ along a line through the origin and perpendicular to
$\hat{\bf a}$.  If we let $D_{\perp}$ and $D_{\parallel}$ be the
distances between the turning points aligned respectively
perpendicular and parallel to $\hat{\bf a}$ then the aspect ratio of
the boundary is
\begin{equation}				
D_{\perp}/D_{\parallel}=J(0)/J(1)\mbox{.}
\label{ratio}
\end{equation}
If $J(1) < J(0) < 0$ then $D_{\perp}/D_{\parallel} > 1 $ and the
elongation is in the direction perpendicular to the alignment, as is
observed in convergent extension.  From Fig.\ \ref{anisotropy}a we see that
$J(0)$ corresponds to our previous $J_{ll}$ while $J(1)$ corresponds
to $J_{ss}$.

We have also studied the minimization of the energy functional Eq.\
(\ref{functional}) numerically for the case where $J$ is chosen to be
a gaussian function.  We approximate the boundary curve by a polygon
of at least 100 sides and use an iterative process that moves down the
energy gradient at constant area.  We have started from many initial
configurations, all of which are simple closed polygons.  The final
boundary curve is always the same and with the correct aspect ratio
Eq.\ (\ref{ratio}).  This iterative process could also be viewed as a
model for the dynamics of convergent extension.  Indeed, with the
addition of additive random forces, the method would be essentially a
Langevin dynamics for the evolution of a highly dissipative system.

In conclusion, we can understand convergent extension as an energy
minimization process, provided the cell-cell adhesive energy has a
certain kind of anisotropy.  This single simple property is sufficient
cause for the cell extension, alignment, and intercalation in the
direction of alignment, that are the characteristics of convergent
extension.  We have characterized the anisotropy required [Eq.\
(\ref{tension}) and Eq.\ (\ref{order})].  We have shown that the final
aspect ratio is independent of the initial configuration and have
shown how it depends on the anisotropy Eq.\ (\ref{ratio}).

We believe our arguments are plausible but realize that they are not
conclusive.  Our modeling neglects many degrees of freedom associated
with cell shape and arrangement, which we think, but cannot prove, are
not crucial.  Our procedure of separately minimizing the bulk and
surface energies is accurate only for a large array of cells.  We do
not see much possibility of doing a lot better by purely analytic
methods.  We have initiated simulations of convergent extension, using
the Potts model and Metropolis dynamics methods of references
\cite{graner:simulation} and \cite{glazier:simulation}, with
anisotropic adhesive energies of the type described in this paper.
The use of anisotropic adhesive energies introduces technical
difficulties in that the energy becomes non-local on the scale of the
size of a cell, which considerably increases the simulation time.
Nevertheless we believe the simulations will eventually substantiate
our conclusions.  Even so, the more difficult question of whether this
explanation of convergent extension is correct remains.  Experiments
that probe the possible anisotropy of cell adhesive energy would be
useful, as would experiments that show the final configuration is
largely independent of the initial configuration.

\acknowledgments

The authors thank Francois Graner for a helpful discussion.


\begin{references}

\bibitem{wolpert:development}
L. Wolpert {\it et~al.}, {\em Principles of Development} (Oxford University
  Press, New York, 1998).

\bibitem{keller:les.houches}
R. Keller and J. Shih,  in {\em Interplay of Genetic and Physical Processes in
  the Development of Biological Form at the Frontier of Physics and Biology},
  {\em Les Houches}, edited by D. Beysens, G. Forgacs, and F. Gail (World
  Scientific, Singapore, 1995), pp.\ 143--153.

\bibitem{steinberg:reconstruction}
M.~S. Steinberg, Science {\bf 141},  401  (1963).

\bibitem{graner:i}
F. Graner, J. theor. Biol. {\bf 164},  455  (1993).

\bibitem{goel:anisotropic}
N.~S. Goel and A.~G. Leith, J. theor. Biol. {\bf 28},  469  (1970).

\bibitem{graner:simulation}
F. Graner and J.~A. Glazier, Phys. Rev. Lett. {\bf 69},  2013  (1992).

\bibitem{glazier:simulation}
J.~A. Glazier and F. Graner, Phys. Rev. E {\bf 47},  2128  (1993).

\bibitem{drasdo:monte.carlo}
D. Drasdo, R. Kree, and J.~S. McCaskill, Phys. Rev. E {\bf 52},  6635  (1995).

\bibitem{weliky:notochord}
M. Weliky, S. Minsuk, R. Keller, and G. Oster, Development {\bf 113},  1231
  (1991).

\bibitem{shih:organizer}
J. Shih and R. Keller, Development {\bf 116},  887  (1992).

\bibitem{wulff:construction} G. Wulff and Z. Kristall, Mineral {\bf
34}, 449 (1901); I.~V. Markov, {\em Crystal Growth for Beginners}
(World Scientific, Singapore, 1995).

\end{references}
\end{document}